\documentstyle[12pt]{article}
\begin{document}

\title{Locally Anisotropic Interactions:\\
II. Torsions and Curvatures of Higher Order Anisotropic Superspaces}
\author{Sergiu I. Vacaru \date{}}
\maketitle

\centerline{\noindent{\em Institute of Applied Physics, Academy of Sciences,}}
\centerline{\noindent{\em 5 Academy str., Chi\c sin\v au 2028,
Republic of Moldova}} \vskip10pt
\centerline{\noindent{ Fax: 011-3732-738149, E-mail: lises@cc.acad.md}}

\begin{abstract}
Torsions, curvatures, structure equations and Bianchi identities for locally
anisotropic superspaces (containing as particular cases different
supersymmetric extensions and prolongations of Riemann, Finsler, Lagrange
and Kaluza--Klein spaces) are investigated.
\end{abstract}

\section{Introduction}

Locally anisotropic superspaces are modeled as vector superbundles
(vs--bund\-les) provided with compatible nonlinear and distinguished
connections (in brief, N-- and d--connections) and metric structures \cite
{vlasg}. Higher order anisotropies are introduced on higher order tangent
superbundles or on corresponding generalizations of vs--bundles called
distinguished vector superbundles bundles, dvs--bundles (see for details
the first companion of this work\cite{vjph1}). The aim of the current paper
is to continue our investigations of basic geometric structures on higher
dimension superspaces with generic local anisotropy and to explore possible
applications in theoretical and mathematical physics \cite{vjmp,vg,vlasg}. In
order to compute torsions, curvatures in dvs--bundles as well to define
components of Bianchi identities and Cartan structure equations we shall
propose a supersymmetric variant of the differential geometric technique
developed in \cite{ma87,ma94,mirata}.

Section 2 contains the basic definitions on nonlinear connection structures
 in locally anisotropic superspaces after \cite{vjph1}. The properties of
 fundamental distinguished geometric objects such as d--tensors and
 d--connections and distinguished torsions and curvatures are correspondingly
  considered in sections 3 and 4. Bianchi and Ricci identities as well Cartan
 structure equations are analyzed in sections 5 and 6. The problem of
compatibility of N--connection, d--connection and metric structures is studied
 in section 7. Section 8 is devoted to the geometry of higher order tangent
 superbundles. In section 9 we present some possible variants of supersymmetric
 extensions of Finsler spaces. Higher order prolongations of Finsler and
 Lagrange superspaces as well higher order Lagrange superspaces are described
 in section 10. Finally, in section 11, we draw some conclusions and
 discussion.

\section{Locally Anisotropic Superspaces: Basic De\-finitions}

Let introduce the necessary definitions and denotations on dvs--bundles \cite
{vjph1,vlasg} (on supermanifolds and superbundles see, for instance, \cite
{dew,rog80,bru}):

Locally a supermanifold (superspace, s--space) $\widetilde{M}$ has some
commuting $\widehat{x}^i$ (indices $i,j,...$ run from $1$ to $n,$ where $n$
is the even dimension of s--space) and anticommuting coordinates $\theta ^{%
\widehat{i}}$(indices $\widehat{i},\widehat{j},...$ $=1,2,...,l,$ where $l$
is the odd dimension of s--space). We shall use coordinates $x^I$ provided
with general indices $I,J,...$ of type $I=(i,\widehat{i}),J=(j,\widehat{j}%
),...$ By $\widehat{{\cal E}}$ we denote a vector superspace (vs--space) of
dimension $\left( m,l\right) .$ With respect to a chosen frame we
parametrize an element $y\in \widehat{{\cal E}}$ as $y=\left( \widehat{y}%
,\zeta \right) =\{y^A=\left( \widehat{y}^a,\zeta ^{\widehat{a}}\right) \},$
where $a=1,2,...,m$ and $\widehat{a}=1,2,...,l.$ Indices of type $A=(a,%
\widehat{a}),B=(b,\widehat{b}),...$ will parametrize objects on vs--spaces.

A distinguished vector superbundle (dvs--bundle)\\ $\widetilde{{\cal E}}%
^{<z>}=(\tilde E^{<z>},\pi ^{<d>},{\cal F}^{<d>},\tilde M),$ with surjective
projection $\pi ^{<z>}:\tilde E^{<z>}\rightarrow \tilde M,$ where $\tilde M$
and $\tilde E^{<z>}$ are respectively base and total s--spaces and the
dv--space ${\cal F}^{<z>}$ is the standard fibre, can be defined \cite{vjph1}
in a usual manner (see \cite{bru,cia,bar,vol}, on vector superbundles, and
\cite{len,ma94,ma87} on vector bundles).

The typical fibre ${\cal F}^{<z>}$ is a distinguished vector superspace
(dvs--space) of dimension $(m,l)$ constructed as an oriented direct sum $%
{\cal F}^{<z>}={\cal F}_{(1)}\oplus {\cal F}_{(2)}\oplus ...\oplus {\cal F}%
_{(z)}$ of vs--spaces ${\cal F}_{(p)},\dim {\cal F}_{(p)}=(m_{(p)},l_{(p)}),$
where $(p)=(1),(2),...,(z),$ \\  $\sum_{p=1}^{p=z}m_{(p)}=m,%
\sum_{p=1}^{p=z}l_{(p)}=l.$

Coordinates on ${\cal F}^{<p>}$ are denoted as
$$
{y}^{<p>}=(y_{(1)},y_{(2)},...,y_{(p)})=({\hat y}_{(1)},{\zeta }_{(1)},{\hat
y}_{(2)},{\zeta }_{(2)},...,{\hat y}_{(p)},{\zeta }_{(p)})=
$$
$$
\{y^{<A>}=({{\hat y}^{<a>}},{\zeta }^{<\hat a>})=({{\hat y}^{<a>}},y^{<\hat
a>})\},
$$
where bracketed indices are correspondingly split on ${\cal F}_{(p)}$%
--components:
$$
<A>=\left( A_{(1)},A_{(2)},...,A_{(p)}\right)
,\\  <a>=(a_{(1)},a_{(2)},...,a_{(p)})
$$
$$
\mbox{ and }<\widehat{a}>=(\widehat{a}_{(1)}\widehat{a}_{(2)},...,\widehat{a}%
_{(p)}),
$$
We shall also write indices in a more simplified form, $<A>=\left(
A_1,A_2,...,A_p\right) ,$\\
 $ <a>=(a_1,a_2,...,a_p)$ and $<\widehat{a}>=(\widehat{%
a}_1,\widehat{a}_2,...,\widehat{a}_p)$ if this will give not rise to
ambiguities.

A dvs--bundle $\widetilde{{\cal E}}^{<z>}$ can be considered as an oriented
set of vs--bundles $\pi ^{<p>}:\tilde E^{<p>}\rightarrow \tilde E^{<p-1>}$
(with typical fibers ${\cal F}^{<p>},p=1,2,...,z);$ $\tilde E^{<0>}=\tilde
M. $ The index $z~(p)$ denotes the total (intermediate) numbers of
consequent vs--bundle coverings of $\tilde M.$ Local coordinates on $%
\widetilde{{\cal E}}^{<p>}$ are parametrized in this manner:
$$
u_{(p)}=(x,y_{<p>})=(x,y_{(1)},y_{(2)},...,y_{(p)})=
({\hat x},{\theta },{\hat y}_{<p>},{\zeta }_{<p>})=$$
$$({\hat x},{\theta },{\hat
y}_{(1)},{\zeta }_{(1)},{\hat y}_{(2)},{\zeta }_{(2)},...,{\hat y}_{(p)},{%
\zeta }_{(p)})=\{u^{<\alpha >}=(x^I,y^{<A>})=({{\hat x}^i},{\theta }^{\hat i},{{\hat y}%
^{<a>}},{\zeta }^{<\hat a>})$$
$$=({{\hat x}^i},x^{\hat i},{{\hat y}^{<a>}},y^{<\hat a>})\}=
(x^I, y^{A_1}, y^{A_2},..., y^{A_p})=...
$$

The coordinate transforms on dvs--bundles\\ $\{u^{<\alpha
>}=(x^I,y^{<A>})\}\rightarrow \{u^{<\alpha ^{\prime }>}=(x^{I^{\prime
}}=x^{I^{\prime }}(x^I),~y^{<A^{\prime }>}=K_{<A>}^{<A^{\prime }>}y^{<A>})\}$
are given by recurrent maps:%
$$
x^{I^{\prime }}=x^{I^{\prime }}({x^I}),{\quad }srank({\frac{{\partial }%
x^{I^{\prime }}}{{\partial }x^I}})=(n,k),\eqno(1)
$$
$$
y_{(1)}^{A_1^{\prime }}=K_{A_1}^{A_1^{\prime }}(x){y}%
_{(1)}^{A_1},...,y_{(p)}^{A_p^{\prime }}=K_{A_p}^{A_p^{\prime }}(u_{(p-1)}){y%
}_{(p)}^{A_p},...,y_{(z)}^{A_z^{\prime }}=K_{A_z}^{A_z^{\prime }}(u_{(z-1)}){%
y}_{(z)}^{A_z},
$$
where $K_{A_p}^{A_p^{\prime }}(u_{(p-1)}){\in }G(m_{(p)},l_{(p)},\Lambda ),
 \Lambda $ is the Grassman algebra (provided with both structures of a Banach
algebra and Euclidean topological space) used for construction of our
superspaces.
A nonlinear connection (N--connection) in a dvs--bundle $\widetilde{{\cal E}}%
^{<z>}$ can be introduced \cite{vjph1} as a regular supersymmetric
distribution $N\left( \widetilde{{\cal E}}^{<z>}\right) $ (horizontal
distribution being supplementary to the vertical s-distribution $V\left(
\widetilde{{\cal E}}^{<z>}\right) )$ determined by maps $N:u\in \widetilde{%
{\cal E}}^{<z>}\rightarrow N(u)\subset T_u\left( \widetilde{{\cal E}}%
^{<z>}\right) $ for which one holds the Whitney sum:
$$
T\left( \widetilde{{\cal E}}^{<z>}\right) =N\left( \widetilde{{\cal E}}%
^{<z>}\right) \oplus V\left( \widetilde{{\cal E}}^{<z>}\right) .\eqno(2)
$$
Locally a N--connection in $\widetilde{{\cal E}}^{<z>}$ is given by its
coefficients
$$
N_{(01)I}^{A_1}(u),(N_{(02)I}^{A_2}(u),N_{(12)A_1}^{A_2}(u)),...,
(N_{(0p)I}^{A_p}(u),N_{(1p)A_1}^{A_p}(u),...N_{(p-1p)A_{p-1}}^{A_p}(u)),...,
$$
$$
(N_{(0z)I}^{A_z}(u),N_{(1z)A_1}^{A_z}(u),...,N_{(pz)A_p}^{A_z}(u),...,
N_{(z-1z)A_{z-1}}^{A_z}(u)).
$$

If a N--connection structure is defined we must extend the operation of
partial derivation in this manner:
$$
{\bf \delta }_{\bullet }{\bf =}\widehat{{\bf N}}(u)\times {\bf \partial }%
_{\bullet },\eqno(3)
$$
where column matrices
$$
{\bf \delta }_{\bullet }{\bf =}\delta _{<\alpha >}=\left(
\begin{array}{c}
\delta _I \\
\delta _{A_1} \\
\delta _{A_2} \\
... \\
\delta _{A_z}
\end{array}
\right) =\left(
\begin{array}{c}
\frac \delta {\partial x^I} \\
\frac \delta {\partial y_{(1)}^{A_1}} \\
\frac \delta {\partial y_{(2)}^{A_2}} \\
... \\
\frac \delta {\partial y_{(z)}^{A_z}}
\end{array}
\right) \mbox{and}~{\bf \partial }_{\bullet }{\bf =}\partial _{<\alpha
>}=\left(
\begin{array}{c}
\partial _I \\
\partial _{A_1} \\
\partial _{A_2} \\
... \\
\partial _{A_z}
\end{array}
\right) =\left(
\begin{array}{c}
\frac \partial {\partial x^I} \\
\frac \partial {\partial y_{(1)}^{A_1}} \\
\frac \partial {\partial y_{(2)}^{A_2}} \\
... \\
\frac \partial {\partial y_{(z)}^{A_z}}
\end{array}
\right)
$$
defines respectively a locally adapted basis (frame) and a usual coordinate
frame and matrix
$$
\widehat{{\bf N}}{\bf =}\left(
\begin{array}{ccccc}
1 & -N_I^{A_1} & -N_I^{A_2} & ... & -N_I^{A_z} \\
0 & 1 & -N_{A_1}^{A_2} & ... & -N_{A_1}^{A_z} \\
0 & 0 & 1 & ... & -N_{A_2}^{A_z} \\
... & ... & ... & ... & ... \\
0 & 0 & 0 & ... & 1
\end{array}
\right)
$$
is constructed by using components of N--connection.

The dual to (3) bases are denoted correspondingly as
$$
{\bf \delta }^{\bullet }={\bf d}^{\bullet }\times {\bf M}(u),\eqno(4)
$$
where
$$
{\bf \delta }^{\bullet }=\left(
\begin{array}{ccccc}
\delta x^I & \delta y^{A_1} & \delta y^{A_2} & ... & \delta y^{A_z}
\end{array}
\right) ,~{\bf d}^{\bullet }=\left(
\begin{array}{ccccc}
dx^I & dy^{A_1} & \delta y^{A_2} & ... & \delta y^{A_z}
\end{array}
\right)
$$
and matrix
$$
{\bf M=}\left(
\begin{array}{ccccc}
1 & M_{(1)I}^{A_1} & M_{(2)I}^{A_2} & ... & M_{(z)I}^{A_z} \\
0 & 1 & M_{(2)A_1}^{A_2} & ... & M_{(z)A_1}^{A_z} \\
0 & 0 & 1 & ... & M_{(z)A_2}^{A_z} \\
... & ... & ... & ... & ... \\
0 & 0 & 0 & ... & 1
\end{array}
\right) ,
$$
defining the dual components of the N--connection, is s--inverse to matrix $%
{\bf N}$ from (3).

With respect to coordinate changes (1) the la-base (3) transforms as
$$
\frac \delta {\partial x^I}=\frac{{\partial }x^{I^{\prime }}}{{\partial }x^I}%
\frac \delta {\partial x^{I^{\prime }}},~\frac \delta {\partial
y_{(p)}^{A_p}}=K_{A_p}^{A_p^{\prime }}\frac \delta {\partial
y_{(p)}^{A_p^{\prime }}}, \forall p=1,2,...,z, \eqno(5)
$$
where $K^{I^{\prime}}_I = \frac{{\partial }x^{I^{\prime}}}{{\partial }x^I}.$

We obtain a supersymmetric generalization of the Miron--Atanasiu \cite
{mirata} osculator bundle $\left( Osc^z\tilde M,\pi ,\tilde M\right) $ if
the fiber space is taken to be a direct sum of $z$ vector s-spaces of the
same dimension $\dim {\cal F}=\dim \widetilde{M},$ i.e. ${\cal F}^{<d>}=%
{\cal F}\oplus {\cal F}\oplus ...\oplus {\cal F}.$ For $z=1$ the $Osc^1%
\widetilde{M}$ is the tangent s-bundle $T\widetilde{M}.$ We also note that
for the osculator s-bundle $\left( Osc^z\tilde M,\pi ,\tilde M\right) $ can
be defined an additional s--tangent structure
$
J:\Xi \left( Osc^z\tilde M\right) \rightarrow \Xi \left( Osc^z\tilde
M\right)
$
defined as
$$
\frac \delta {\partial y_{(1)}^I}=J\left( \frac \delta {\partial x^I}\right)
,...,\frac \delta {\partial y_{(z-1)}^I}=J\left( \frac \delta {\partial
y_{(z-2)}^I}\right) ,\frac \partial {\partial y_{(z)}^I}=J\left( \frac
\delta {\partial y_{(z-1)}^I}\right). \eqno(6)
$$

In order to simplify considerations in this work we shall consider only
locally trivial vector bundles.

\section{Distinguished Tensors and Connections in DVS--Bundles}

By using adapted bases (3) and (4) one introduces the algebra $DT({\tilde {%
{\cal E}}}^{<z>})$ of distinguished tensor s-fields (ds-fields, ds-tensors,
ds-objects) on $\tilde {{\cal E}}^{<z>},{\quad }{\cal T}={\cal T}%
_{qq_1q_2...q_z}^{pp_1p_{2....}p_z},$ which is equivalent to the tensor
algebra of vs-bundle ${\pi }_{hv_1v_2...v_z}:H\tilde {{\cal E}}^{<z>}{\oplus
}V_1\tilde {{\cal E}}^{<z>}{\oplus }V_2\tilde {{\cal E}}^{<z>}\oplus ...{%
\oplus }V_s\tilde {{\cal E}}^{<z>}{\to }\tilde {{\cal E}}^{<z>}.$ An element
$Q{\in {\cal T}}_{qq_1q_2...q_z}^{pp_1p_{2....}p_z},$ , ds-field of type $%
\left(
\begin{array}{ccccc}
p & p_1 & p_2 & ... & p_z \\
q & q_1 & q_2 & ... & q_z
\end{array}
\right) ,$ can be written in local form as
$$
Q={Q}_{{J_1}{\dots }{J_q}{B_1}{\dots }{B}_{q_1}{C}%
_1...C_{q_2}...F_1...F_{q_s}}^{{I_1}{\dots }{I_p}{A_1}{\dots }{A}%
_{p_1}E_1...E_{p_2}...D_1....D_{p_s}}{(u)}{{\delta }_{I_1}}\otimes {\dots }%
\otimes {{\delta }_{I_p}}\otimes {d^{J_1}}\otimes {\dots }\otimes {d^{J_q}}%
\otimes
$$
$$
{{\delta }_{A_1}}\otimes {\dots }\otimes {\delta }_{A_{p_1}}\otimes {{%
\delta }^{B_1}{\delta }^{B_1}}\otimes {\dots }\otimes {\delta }%
^{B_{q_1}}\otimes {{\delta }_{E_1}}\otimes {\dots }\otimes {\delta }%
_{E_{p_2}}\otimes {{\delta }^{C_1}}\otimes {\dots }\otimes {\delta }%
^{C_{q_2}}\otimes ...
$$
$$
\otimes {{\delta }_{D_1}}\otimes {\dots }\otimes {\delta }%
_{D_{p_z}}\otimes {{\delta }^{F_1}}\otimes {\dots }\otimes {\delta }%
^{F_{qz}}.\eqno(7)
$$

In addition to ds-tensors we can introduce ds-objects with various s-group
and coordinate transforms adapted to a global splitting (2).

A linear distinguished connection, d- connection, in dvs-bundle $\tilde {%
{\cal E}}^{<z>}$ is a linear connection $D$ on $\tilde {{\cal E}}^{<z>}$
which preserves by parallelism the horizontal and vertical distributions in $%
\tilde {{\cal E}}^{<z>}$.

By a linear connection of a s-manifold we understand a linear connection in
its tangent bundle.

Let denote by $\Xi (\tilde M)$ and $\Xi (\tilde {{\cal E}}^{<p>}),$
respectively, the modules of vector fields on s-manifold $\tilde M$ and
dvs-bundle $\tilde {{\cal E}}^{<p>}$ and by ${\cal F}{(\tilde M)}$ and $%
{\cal F}{(\tilde {{\cal E}}}^{<p>}{)},$ respectively, the s-modules of
functions on $\tilde M$ and on $\tilde {{\cal E}}^{<p>}.$

It is clear that for a given global splitting into horizontal and verticals
s-subbund\-les (2) we can associate operators of horizontal and vertical
covariant derivations (h- and v-derivations, denoted respectively as $D^{(h)}
$ and $D^{(v_1v_2...v_z)}$) with properties:
$$
{D_X}Y=(XD)Y={D_{hX}}Y+{D}_{v_1X}Y+{D}_{v_2X}Y+...+{D}_{v_zX}Y,
$$
where
$$
D_X^{(h)}{Y}=D_{hX}{Y},{\quad }D_X^{(h)}f=(hX)f
$$
and
$$
D_X^{(v_p)}{Y}=D_{v_pX}{Y},{\quad }D_X^{(v_p)}f=(v_pX)f,~(p=1,...,z)
$$
for every $f\in {\cal F}(\tilde M)$ with decomposition of vectors $X,Y\in {%
\Xi }(\tilde {{\cal E}}^{<z>})$ into horizontal and vertical parts, $%
X=hX+v_1X+....+v_zX{\quad }$ and ${\quad }Y=hY+v_1Y+...+v_zY.$

The local coefficients of a d- connection $D$ in $\tilde {{\cal E}}^{<z>}$
with respect to the local adapted frame (3) separate into corresponding
distinguished groups. We introduce horizontal local coefficients\\ $%
(L_{JK}^I,L_{<B>K}^{<A>})=({L^I}_{JK}(u),{L}_{B_1K}^{A_1}{(}u),{L}%
_{B_2K}^{A_2}{(}u),...,{L}_{B_zK}^{A_z}{(}u))$ of $D^{(h)}$ such that
$$
D_{({\frac \delta {\delta x^K}})}^{(h)}{\frac \delta {\delta x^J}}={L^I}%
_{JK}(u){\frac \delta {\delta x^I}},D_{({\frac \delta {\delta x^K}}%
)}^{(h)}\frac \delta {\delta y_{(p)}^{B_p}}={L}_{B_pK}^{A_p}{(u)}\frac
\delta {\delta y_{(p)}^{A_p}},(p=1,...,z),
$$
$$
D_{({\frac \delta {\delta x^k}})}^{(h)}q={\frac{{\delta q}}{\delta x^K}},
$$
and $p$--vertical local coefficients\\ $(C_{J<C>}^I,C_{<B><C>}^{<A>})=({C}%
_{JC_p}^I(u),{C}_{B_1C_p}^{A_1}(u),{C}_{B_2C_p}^{A_2}(u),...,{C}%
_{B_zC_p}^{A_z}(u))$ $(p=1,...,z)$ such that
$$
D_{(\frac \delta {\delta y^{C_p}})}^{(v_p)}{\frac \delta {\delta x^J}}={C}%
_{JC_p}^I(u){\frac \delta {\delta x^I}},D_{(\frac \delta {\delta
y^{C_p}})}^{(v_p)}\frac \delta {\delta y_{(f)}^{B_f}}={C}_{B_fC_p}^{A_f}%
\frac \delta {\delta y_{(f)}^{A_f}},D_{(\frac \delta {\delta
y^{C_p}})}^{(v_p)}q=\frac{\delta q}{\partial y^{C_p}},
$$
where $q\in {\cal F}(\tilde {{\cal E}}^{<z>}),$ $f=1,...,z.$

The covariant ds-derivation along vector $X=X^I{\frac \delta {\partial x^I}}%
+Y^{A_1}\frac \delta {\partial y^{A_1}}+...+Y^{A_z}\frac \partial {\partial
y^{A_z}}$ of a ds-tensor field $Q,$ for instance, of type $\left(
\begin{array}{cc}
p & p_r \\
q & q_r
\end{array}
\right) ,1\leq r\leq z,{\quad }$ see (7), can be written as
$$
{D_X}Q=D_X^{(h)}Q+D_X^{(v_1)}Q+...+D_X^{(v_z)}Q,
$$
where h-covariant derivative is defined as
$$
D_X^{(h)}Q=X^KQ_{JB_r{\mid }K}^{IA_r}{\delta }_I{\otimes \delta }_{A_r}{%
\otimes }d^J{\otimes }{\delta }^{B_r},
$$
with components

$$
Q_{JB_r{\mid }K}^{IA_r}={\frac{\delta Q_{JB_r}^{IA_r}}{\partial x^K}}+{L^I}%
_{HK}Q_{JB_R}^{HA_r}+{L}_{C_iK}^{A_r}Q_{JB_i}^{IC_r}-{L^H}%
_{JK}Q_{HB_r}^{IA_r}-{L}_{B_rK}^{C_r}Q_{JC_r}^{IA_r},
$$
and v$_p$-covariant derivatives defined as
$$
{D}_X^{(v_p)}Q={X}^{C_p}{Q}_{JB_r\perp C_p}^{IA_r}{\delta }_I{\otimes }{%
\delta }_{A_r}{\otimes }{\delta}^I{\otimes }{\delta }^{B_r},
$$
with components
$$
{Q}_{JB_r\perp C_p}^{IA_r}=\frac{\delta Q_{JB_R}^{IA_r}}{\partial y^{C_p}}+{%
C^I}_{HC_p}Q_{JB_R}^{HA_r}+{C}_{F_rC_p}^{A_r}Q_{JB_R}^{IF_r}-{C}%
_{JC_p}^HQ_{HF_R}^{IA_r}-{C}_{B_rC_p}^{F_r}Q_{JF_R}^{IA_r}..
$$

The above presented formulas show that $D{\Gamma }=(L,...,{L}_{(p)},...,{C}%
,...,C_{(p)},...)$ are the local coefficients of the d-connection $D$ with
respect to the local frame $({\frac \delta {\delta x^I}},{\frac \delta
{\partial y^{<A>}}}).$ If a change (1) of local coordinates on
$\tilde {{\cal E}%
}^{<z>}$ is performed we have the following transformation laws of the local
coefficients of a d-connect\-i\-on:

$$
{L^{I^{\prime }}}_{J^{\prime }M^{\prime }}={\frac{\partial x^{I^{\prime }}}{%
\partial x^I}}{\frac{\partial x^J}{\partial x^{J^{\prime }}}}{\frac{\partial
x^M}{\partial x^{M^{\prime }}}}{L^I}_{JM}+{\frac{\partial x^{I^{\prime }}}{%
\partial x^M}}{\frac{{\partial }^2x^M}{{{\partial x^{J^{\prime }}}{\partial
x^{M^{\prime }}}}}},\eqno(8)
$$
$$
{L}_{(f)B_f^{\prime }M^{\prime }}^{A_f^{\prime }}=K_{A_f}^{A_f^{\prime
}}K_{B_f^{\prime }}^{B_f}{\frac{\partial x^M}{\partial x^{M^{\prime }}}L}%
_{(f)B_fM}^{A_f}+K_{C_f}^{A_f^{\prime }}\frac{\partial K_{B_f^{\prime
}}^{C_f}}{\partial x^{M^{\prime }}},
$$
$$
................
$$
$$
{C_{(p)J^{\prime }C_p^{\prime }}^{I^{\prime }}}={\frac{\partial x^{I^{\prime
}}}{\partial x^I}}{\frac{\partial x^J}{\partial x^{J^{\prime }}}}%
K_{C_p}^{C_p^{\prime }}{C_{(p)JC_p}^I},...,{C_{B_f^{\prime }C_p^{\prime
}}^{A_f^{\prime }}}=K_{A_f}^{A_f^{\prime }}{K}_{B_f^{\prime
}}^{B_f}K_{C_p^{\prime }}^{C_p}{C_{B_fC_p}^{A_f},...}.
$$
As in the usual case of tensor calculus on locally isotropic spaces the
transformation laws (8) for d-connections differ from those for ds-tensors,
which are written (for instance, we consider transformation laws for
ds-tensor (7)) as
$$
{Q}_{{J}^{\prime }{_1}{\dots }{J}^{\prime }{_q}{B}^{\prime }{_1}{\dots }{B}%
_{q_1}^{\prime }{C}_1^{\prime }...C_{q_2}^{\prime }...F_1^{\prime
}...F_{q_s}^{\prime }}^{{I}^{\prime }{_1}{\dots }{I}^{\prime }{_p}{A}%
^{\prime }{_1}{\dots }{A}_{p_1}^{\prime }E_1^{\prime }...E_{p_2}^{\prime
}...D_1^{\prime }....D_{p_s}^{\prime }}=
$$
$$
{\frac{\partial x^{I_1^{\prime }}}{\partial x^{I_1}}{\frac{\partial x^{J_1}}{%
\partial x^{J_1^{\prime }}}}\dots }K_{A_1}^{A_1^{\prime }}K_{B_1^{\prime
}}^{B_1}{\dots K}_{D_{p_s}}^{D_{p_s}^{\prime }}{K}_{F_{p_s}^{\prime
}}^{F_{p_s}}{Q}_{{J_1}{\dots }{J_q}{B_1}{\dots }{B}_{q_1}{C}%
_1...C_{q_2}...F_1...F_{q_s}}^{{I_1}{\dots }{I_p}{A_1}{\dots }{A}%
_{p_1}E_1...E_{p_2}...D_1....D_{p_s}}.
$$

To obtain local formulas on usual higher order anisotropic spaces we have to
restrict us with even components of geometric objects by changing, formally,
capital indices $(I,J,K,...)$ into $(i,j,k,a,..)$ and s-derivation and
s-commutation rules into those for real number fields on usual manifolds. We
shall consider various applications in the theoretical and mathematical
physics of the differential geometry of distinguished vector bundles in our
further works.

\section{ Torsions and Curvatures of D--Connections}

Let $\tilde {{\cal E}}^{<z>}$ be a dvs--bundle endowed with N-connection and
d-connec\-ti\-on structures. The torsion of a d-connection is introduced as
$$
T(X,Y)=[X,DY\}-[X,Y\},{\quad }X,Y{\subset }{\Xi }{(\tilde M)}.
$$
where $[...\}$ is the s--commutator. One holds the following invariant
decomposition (by using h-- and v--projections associated to N):
$$
T(X,Y)=T(hX,hY)+T(hX,v_1Y)+T(v_1X,hX)+T(v_1X,v_1Y)+...
$$
$$
+T(v_{p-1}X,v_{p-1}Y)+T(v_{p-1}X,v_pY)+T(v_pX,v_{p-1}X)+T(v_pX,v_pY)+...
$$
$$
+T(v_{z-1}X,v_{z-1}Y)+T(v_{z-1}X,v_zY)+T(v_zX,v_{z-1}X)+T(v_zX,v_zY).
$$
Taking into ac\-count the skew\-su\-per\-sym\-me\-try of $T$ and the
equa\-tions
$$
h[v_pX,v_pY\}=0,...,v_f[v_pX,v_pY\}=0,f\neq p,
$$
we can verify that the torsion of a d-connection is completely determined by
the following ds-tensor fields:
$$
hT(hX,hY)=[X(D^{(h)}{h})Y\}-h[hX,hY\},...,
$$
$$
v_pT(hX,hY)=-v_p[hX,hY\},...,
$$
$$
hT(hX,v_pY)=-D_Y^{(v_p)}{hX}-h[hX,v_pY\},...,
$$
$$
v_pT(hX,v_pY)=D_X^{(h)}{v}_p{Y}-v_p[hX,v_pY\},...,
$$
$$
v_fT(v_fX,v_fY)=[X(D^{(v_f)}{v}_f)Y\}-v_f[v_fX,v_fY\},...,
$$
$$
v_pT(v_fX,v_fY)=-v_p[v_fX,v_fY\},...,
$$
$$
v_fT(v_fX,v_pY)=-D_Y^{(v_p)}{v}_f{X}-v_f[v_fX,v_pY\},...,
$$
$$
v_pT(v_fX,v_pY)=D_X^{(v_f)}{v}_p{Y}-v_p[v_fX,v_pY\},...,f<p,
$$
$$
v_{z-1}T(v_{z-1}X,v_{x-1}Y)=[X(D^{(v_{z-1})}{v}_{z-1})Y%
\}-v_{z-1}[v_{z-1}X,v_{z-1}Y\},...,
$$
$$
v_zT(v_{z-1}X,v_{z-1}Y)=-v_z[v_{z-1}X,v_{z-1}Y\},
$$
$$
v_{z-1}T(v_{z-1}X,v_zY)=-D_Y^{(v_z)}{v}_{z-1}{X}-v_{z-1}[v_{z-1}X,v_zY%
\},...,
$$
$$
v_zT(v_{z-1}X,v_zY)=D_X^{(v_{z-1})}{v}_z{Y}-v_z[v_{z-1}X,v_zY\}.
$$
where $X,Y\in {{\Xi }(\tilde {{\cal E}}^{<z>})}.$ In order to get the local
form of the ds-tensor fields which determine the torsion of d-connection $D{%
\Gamma }$ (the torsions of $D{\Gamma }$) we use equations
$$
[{\frac \delta {\partial x^J}},{\frac \delta {\partial x^K}}\}={R}%
_{JK}^{<A>}\frac \delta {\partial y^{<A>}},~[{\frac \delta {\partial x^J}},{%
\frac \delta {\partial y^{<B>}}}\}={\frac{\delta {N_J^{<A>}}}{\partial
y^{<B>}}}{\frac \delta {\partial y^{<A>}}},
$$
where
$$
{R}_{JK}^{<A>}=\frac{\delta N_J^{<A>}}{\partial x^K}-{(-)}^{\mid {KJ}\mid }%
\frac{\delta N_K^{<A>}}{\partial x^J},
$$
$$
{(-)}^{{\mid K\mid }{\mid J\mid }}={(-)}^{\mid {KJ}\mid }= {(-1)}^{\mid {KJ}%
\mid },%
$$
$|K|=0 (1)$ for even (odd) components, and introduce notations
$$
hT({\frac \delta {\delta x^K}},{\frac \delta {\delta x^J}})={T^I}_{JK}{\frac
\delta {\delta x^I}},\
v_1T{({\frac \delta {\delta x^K}},{\frac \delta {\delta x^J}})}={\tilde T}%
_{KJ}^{A_1}\frac \delta {\partial y^{A_1}}, \eqno(9)
$$
$$
hT({\frac \delta {\partial y^{<A>}}},{\frac \delta {\partial x^J}})={\tilde P%
}_{J<A>}^I{\frac \delta {\delta x^I}},...,\
v_pT(\frac \delta {\partial y^{B_p}},{\frac \delta {\delta x^J}})={P}%
_{JB_p}^{<A>}{\frac \delta {\partial y^{<A>}}},...,
$$
$$
v_pT(\frac \delta {\partial y^{C_p}},\frac \delta {\partial y^{B_f}})={S}%
_{B_fC_p}^{<A>}{\frac \delta {\partial y^{<A>}}}.
$$

Now we can compute the local components of the torsions, introduced in (9),
with respect to a la-frame of a d-connection\\ $D{\Gamma }=(L,...,{L}%
_{(p)},...,{C},...,C_{(p)},...)$ $:$
$$
{T^I}_{JK}={L^I}_{JK}-{(-)}^{\mid {JK}\mid }{L^I}_{KJ},\
{T}^{<A>}_{JK}={R^{<A>}}_{JK}, \eqno(10)$$
$${P^I}_{J<B>}={C^I}_{J<B>},\
{P^{<A>}}_{J<B>}={\frac{\delta {N_J^{<A>}}}{\partial y^{<B>}}}-{L^{<A>}}%
_{<B>J},
$$
$$
{S^{<A>}}_{<B><C>}={C^{<A>}}_{<B><C>}-{(-)}^{\mid <B><C>\mid }{C^{<A>}}%
_{<C><B>}.
$$
The even and odd components of torsions (10) can be specified in explicit
form by using decompositions of indices into even and odd parts $(I=(i,{\hat
i}),J=(j,{\hat j}),..)$, for instance,
$$
{T^i}_{jk}={L^i}_{jk}-{L^i}_{kj},{\quad }{T^i}_{j{\hat k}}={L^i}_{j{\hat k}}+%
{L^i}_{{\hat k}j},
$$
$$
{T^{\hat i}}_{jk}={L^{\hat i}}_{jk}-{L^{\hat i}}_{kj},{\dots },
$$
and so on (see \cite{ma87,ma94} and \cite{vg,vjmp} explicit formulas on
torsions on a la-space $M$, one omits ''tilde'' for usual manifolds and
bundles).

Another important characteristic of a d-connection $D\Gamma $ is its
curvature:
$$
R(X,Y)Z=D_{[X}D_{Y\}}-D_{[X,Y\}}Z,
$$
where $X,Y,Z\in {\Xi }(\tilde E^{<z>}).$ From the properties of h- and
v-projections it follows that
$$
v_pR(X,Y)hZ=0,...,hR(X,Y)v_pZ=0,v_fR(X,Y)v_pZ=0,f\neq p,\eqno(11)
$$
and
$$
R(X,Y)Z=hR(X,Y)hZ+v_1R(X,Y)v_1Z+...+v_zR(X,Y)v_zZ,
$$
where $X,Y,Z\in {\Xi }(\tilde E^{<z>}).$ Tak\-ing into ac\-count
prop\-er\-ties (11) and the equa\-tions
$$
R(X,Y)=-{(-)}^{\mid XY\mid }R(Y,X)
$$
we prove that the curvature of a d-connection $D$ in the total space of a
vs-bundle $\tilde E^{<z>}$ is completely determined by the following
ds-tensor fields:
$$
R(hX,hY)hZ=({D_{[X}^{(h)}}D_{Y\}}^{(h)}-D_{[hX,hY\}}^{(h)}\eqno(12)
$$
$$
-D_{[hX,hY\}}^{(v_1)}-...D_{[hX,hY\}}^{(v_{z-1})}-D_{[hX,hY\}}^{(v_z)})hZ,
$$
$$
R(hX,hY)v_pZ=(D_{[X}^{(h)}D_{Y\}}^{(h)}-D_{[hX,hY\}}^{(h)}-
$$
$$
D_{[hX,hY\}}^{(v_1)})v_pZ-...-D_{[hX,hY\}}^{(v_{p-1})})v_pZ-D_{[hX,hY%
\}}^{(v_p)})v_pZ,
$$
$$
R(v_pX,hY)hZ=(D_{[X}^{(v_p)}D_{Y\}}^{(h)}-D_{[v_pX,hY\}}^{(h)}-
$$
$$
D_{[v_pX,hY\}}^{(v_{_1})})hZ-...-D_{[v_pX,hY\}}^{(v_{p-_1})}-D_{[v_pX,hY%
\}}^{(v_p)})hZ,
$$
$$
R(v_fX,hY)v_pZ=(D_{[X}^{(v_f)}D_{Y\}}^{(h)}-D_{[v_fX,hY\}}^{(h)}-
$$
$$
D_{[v_fX,hY\}}^{(v_1)})v_1Z-...-D_{[v_fX,hY%
\}}^{(v_{p-1})})v_{p-1}Z-D_{[v_fX,hY\}}^{(v_p)})v_pZ,
$$
$$
R(v_fX,v_pY)hZ=(D_{[X}^{(v_f)}D_{Y\}}^{(v_p)}-D_{[v_fX,v_pY%
\}}^{(v_1)})hZ-...
$$
$$
-D_{[v_fX,v_pY\}}^{(v_{z-1})}-D_{[v_fX,v_pY\}}^{(v_z)})hZ,
$$
$$
R(v_fX,v_qY)v_pZ=(D_{[X}^{(v_f)}D_{Y\}}^{(v_q)}-D_{[v_fX,v_qY%
\}}^{(v_1)})v_1Z-...
$$
$$
-D_{[v_fX,v_qY\}}^{(v_{p-1})})v_{p-1}Z-D_{[v_fX,v_qY\}}^{(v_p)})v_pZ,
$$
where
$$
{D_{[X}^{(h)}}{D_{Y\}}^{(h)}}={D_X^{(h)}}{D_Y^{(h)}}-{{(-)}^{\mid XY\mid }}{%
D_Y^{(h)}}{D_X^{(h)}},~
$$
$$
{D_{[X}^{(h)}}{D}_{Y\}}^{(v_p)}={D_X^{(h)}}{D}_Y^{(v_p)}-{(-)}^{|Xv_pY|}{D}%
_Y^{(v_p)}{D_X^{(h)},}
$$
$$
{D}_{[X}^{(v_p)}{D_{Y\}}^{(h)}}={D}_X^{(v_p)}{D_Y^{(h)}}-{(-)}^{\mid
v_pXY\mid }{D_Y^{(h)}}{D}_X^{(v_p)},~
$$
$$
~{D}_{[X}^{(v_f)}{D}_{Y\}}^{(v_p)}={D}_X^{(v_f)}{D}_Y^{(v_p)}-{(-)}%
^{|v_fXv_pY|}{D}_Y^{(v_p)}{D}_X^{(v_f)}.
$$
The local components of the ds-tensor fields (12) are introduced as follows:
$$
R({{\delta }_K},{{\delta }_J}){{\delta }_H}={{{R_H}^I}_{JK}}{{\delta }_I},{~}%
R({{\delta }_K},{{\delta }_J})\delta {_{<B>}}={R}_{<B>.JK}^{.<A>}\delta
_{<A>},\eqno(13)
$$
$$
R({\delta _{<C>}},{{\delta }_K}){{\delta }_J}=P_{JK<C>}^I{{{\delta }_I},~}%
R\left( \delta _{<C>},\delta _K\right) \delta
_{<B>}=P_{<B>.K<C>}^{.<A>}\delta _{<A>},
$$
$$
R({\delta _{<C>}},{{\delta }_{<B>}}){{\delta }_J}=S_{J.<B><C>}^{.I}{{{\delta
}_I}},R({\delta _{<D>}},{{\delta }_{<C>}}){{\delta }_{<B>}}%
=S_{<B>.<C><D>}^{.<A>}{{{\delta }_{<A>}}}.
$$
Putting the components of a d-connection $D{\Gamma }=(L,...,{L}_{(p)},...,{C}%
,...,C_{(p)},...)$ in (13), by a direct computation, we obtain these locally
adapted components of the curvature (curvatures):
$$
{{R_H}^I}_{JK}={{\delta }_K}{L^I}_{HJ}-{(-)}^{\mid KJ\mid }{{\delta }_J}{L^I}%
_{HK}+
$$
$$
{L^M}_{HJ}{L^I}_{MK}-{(-)}^{\mid KJ\mid }{L^M}_{HK}{L^I}_{MJ}+{C^I}_{H<A>}{%
R^{<A>}}_{JK},
$$
$$
{R_{<B>\cdot JK}^{\cdot <A>}}={{\delta }_K}{L^{<A>}}_{<B>J}-{(-)}^{\mid
KJ\mid }{{\delta }_J}{L^{<A>}}_{<B>K}+
$$
$$
{L^{<C>}}_{<B>J}{L^{<A>}}_{<C>K}-{(-)}^{\mid KJ\mid }{L^{<C>}}_{<B>K}+{%
C^{<A>}}_{<B><C>}{R^{<C>}}_{JK},
$$
$$
{P_{J\cdot K<A>}^{\cdot I}}={\delta _{<A>}}{L^I}_{JK}-{C^I}_{J<A>{\mid }K}+{%
C^I}_{J<B>}{P^{<B>}}_{K<A>},\eqno(14)
$$
$$
{{P_{<B>}}^{<A>}}_{K<C>}={\delta _{<C>}}{L^{<A>}}_{<B>K}-{C^{<A>}}_{<B><C>{%
\mid }K}+{C^{<A>}}_{<B><D>}{P^{<D>}}_{K<C>},
$$
$$
{S_{J\cdot <B><C>}^{\cdot I}}={\delta _{<C>}}{C^I}_{J<B>}-{(-)}^{\mid
<B><C>\mid }{\delta _{<B>}}{C^I}_{J<C>}+
$$
$$
{C^{<H>}}_{J<B>}{C^I}_{<H><C>}-{(-)}^{\mid <B><C>\mid }{C^{<H>}}_{J<C>}{C^I}%
_{<H><B>},
$$
$$
{{S_{<B>}}^{<A>}}_{<C><D>}={\delta _{<D>}}{C^{<A>}}_{<B><C>}-{(-)}^{\mid
<C><D>\mid }{\delta _{<C>}}{C^{<A>}}_{<B><D>}+
$$
$$
{C^{<E>}}_{<B><C>}{C^{<A>}}_{<E><D>}-{(-)}^{\mid <C><D>\mid }{C^{<E>}}%
_{<B><D>}{C^{<A>}}_{<E><C>}.
$$
Even and odd components of curvatures (14) can be written out by splitting
indices into even and odd parts, for instance,
$$
{{R_h}^i}_{jk}={{\delta }_k}{L^i}_{hj}-{{\delta }_j}{{L^i}_{hk}+{{L^m}_{hj}}{%
{L^i}_{mk}}-{{L^m}_{hk}}{{L^i}_{mj}}}+{{C^i}_{h<a>}}{{R^{<a>}}_{jk}},
$$
$$
{{R_h}^i}_{j{\hat k}}={{\delta }_{\hat k}}{L^i}_{hj}+{{\delta }_j}{L^i}_{h{%
\hat k}}+{L^m}_{hj}{L^i}_{m{\hat k}}+{L^m}_{h{\hat k}}{L^i}_{mj}+{C^i}_{h<a>}%
{R^{<a>}}_{j{\hat k}}{\quad },{\dots }.
$$
We omit explicit formulas for even--odd components because we shall not use
them in this work.

\section{Bianchi and Ricci Identities}

The torsions and curvatures of every linear connection $D$ on a vs-bundle $%
\tilde {{\cal E}}^{<z>}$ \\satisfy the following generalized Bianchi
identities:
$$
\sum_{SC}{[(D_X{T})(Y,Z)-R(X,Y)Z+T(T(X,Y),Z)]}=0,
$$
$$
\sum_{SC}{[(D_X{R})(U,Y,Z)+R(T(X,Y)Z)U]}=0,\eqno(15)
$$
where $\sum_{SC}$ means the respective supersymmretric cyclic sum over $%
X,Y,Z $ and $U.$ If $D$ is a d-connection, then by using (11) and
$$
v_p(D_X{R})(U,Y,hZ)=0,{~}h({D_X}R(U,Y,v_pZ)=0,v_f(D_X{R})(U,Y,v_pZ)=0,
$$
the identities (15) become
$$
\sum_{SC}[h({D_X}T)(Y,Z)-hR(X,Y)Z+hT(hT(X,Y),Z)+
$$
$$
hT(v_1T(X,Y),Z)+...+hT(v_zT(X,Y),Z)]=0,
$$
$$
\sum_{SC}{[v}_f{({D_X}T)(Y,Z)}-{v}_f{R(X,Y)Z+}
$$
$$
{v}_f{T(hT(X,Y),Z)+}\sum\limits_{p\geq f}{v}_f{T(v}_p{T(X,Y),Z)]}=0,
$$
$$
\sum_{SC}{[h({D_X}R)(U,Y,Z)+hR(hT(X,Y),Z)U+}
$$
$$
{hR(v}_1{T(X,Y),Z)U+...+{hR(v}_z{T(X,Y),Z)U}]}=0,
$$
$$
\sum_{SC}{[v}_f{({D_X}R)(U,Y,Z)+v}_f{R(hT(X,Y),Z)U+}
$$
$$
\sum\limits_{p\geq f}{v}_f{R(v}_p{T(X,Y),Z)U]}=0.\eqno(16)
$$

In order to get the component form of these identities we insert
correspondingly in (16) these values of triples $(X,Y,Z)$,\ ($=({{\delta }_J}%
,{{\delta }_K},{{\delta }_L}),$ or\\ $({\delta _{<D>}},{\delta _{<C>}},{%
\delta _{<B>}})$), and put successively $U={\delta }_H$ and $U={\delta }%
_{<A>}.$ Taking into account (9),(10) and (12),(13) we obtain:
$$
\sum_{SC[L,K,J\}}[{T^I}_{JK{\mid }H}+{T^M}_{JK}{T^J}_{HM}+{R^{<A>}}_{JK}{C^I}%
_{H<A>}-{{R_J}^I}_{KH}]=0,
$$
$$
\sum_{SC[L,K,J\}}[{{R^{<A>}}_{JK{\mid }H}}+{T^M}_{JK}{R^{<A>}}_{HM}+{R^{<B>}}%
_{JK}{P^{<A>}}_{H<B>}]=0,
$$
{\cal
$$
{C^I}_{J<B>{\mid }K}-{(-)}^{\mid JK\mid }{C^I}_{K<B>{\mid }J}-{T^I}_{JK{\mid
<}B>}+{C^M}_{J<B>}{T^I}_{KM}-
$$
}%
$$
{(-)}^{\mid JK\mid }{C^M}_{K<B>}{T^I}_{JM}+{T^M}_{JK}{C^I}_{M<B>}+{P^{<D>}}%
_{J<B>}{C^I}_{K<D>}
$$
$$
-{(-)}^{\mid KJ\mid }{P^{<D>}}_{K<B>}{C^I}_{J<D>}+{{P_J}^I}_{K<B>}-{(-)}%
^{\mid KJ\mid }{{P_K}^I}_{J<B>}=0,
$$
$$
{P^{<A>}}_{J<B>{\mid }K}-{(-)}^{\mid KJ\mid }{P^{<A>}}_{K<B>{\mid }J}-{%
R^{<A>}}_{JK\perp <B>}+{C^M}_{J<B>}{R^{<A>}}_{KM}-
$$
$$
{(-)}^{\mid KJ\mid }{C^M}_{K<B>}{R^{<A>}}_{JM}+{T^M}_{JK}{P^{<A>}}_{M<B>}+{%
P^{<D>}}_{J<B>}{P^{<A>}}_{K<D>}-
$$
$$
{(-)}^{\mid KJ\mid }{P^{<D>}}_{K<B>}{P^{<A>}}_{J<D>}-{{R^{<D>}}_{JK}}{{%
S^{<A>}}_{B<D>}}+{R_{<B>\cdot JK}^{\cdot <A>}}=0,
$$
$$
{C^I}_{J<B>\perp <C>}-{(-)}^{\mid <B><C>\mid }{C^I}_{J<C>\perp <B>}+{C^M}%
_{J<C>}{C^I}_{M<B>}-
$$
$$
{(-)}^{\mid <B><C>\mid }{C^M}_{J<B>}{C^I}_{M<C>}+{S^{<D>}}_{<B><C>}{C^I}%
_{J<D>}-{S_{J\cdot <B><C>}^{\cdot I}}=0,
$$
$$
{P^{<A>}}_{J<B>\perp <C>}-{(-)}^{\mid <B><C>\mid }{P^{<A>}}_{J<C>\perp <B>}+
$$
$$
{S^{<A>}}_{<B><C>\mid J}+{C^M}_{J<C>}{P^{<A>}}_{M<B>}-
$$
$$
{(-)}^{\mid <B><C>\mid }{C^M}_{J<B>}{P^{<A>}}_{M<C>}+{P^{<D>}}_{J<B>}{S^{<A>}%
}_{<C><D>}-
$$
$$
{(-)}^{\mid <C><B>\mid }{P^{<D>}}_{J<C>}{S^{<A>}}_{<B><D>}+{S^{<D>}}_{<B><C>}%
{P^{<A>}}_{J<D>}+
$$
$$
{{P_{<B>}}^{<A>}}_{J<C>}-{(-)}^{\mid <C><B>\mid }{{P_{<C>}}^{<A>}}_{J<B>}=0,
$$
$$
\sum_{SC[<B>,<C>,<D>\}}[{S^{<A>}}_{<B><C>\perp <D>}+
$$
$$
{S^{<F>}}_{<B><C>}{S^{<A>}}_{<D><F>}-{{S_{<B>}}^{<A>}}_{<C><D>}]=0,
$$
$$
\sum_{SC[H,J,L\}}[{{R_K}^I}_{HJ\mid L}-{T^M}_{HJ}{{R_K}^I}_{LM}-{{R^{<A>}}%
_{HJ}P}{_{K\cdot L<A>}^{\cdot I}}]=0,
$$
$$
\sum_{SC[H,J,L\}}[{R_{<D>\cdot HJ\mid L}^{\cdot <A>}}-{{T^M}_{HJ}R}{%
_{<D>\cdot LM}^{\cdot <A>}}-{{R^{<C>}}_{HJ}}{{{P_{<D>}}^{<A>}}_{L<C>}}]=0,
$$
$$
{P_{K\cdot J<D>\mid L}^{\cdot I}}-{(-)}^{\mid LJ\mid }{P_{K\cdot L<D>\mid
J}^{\cdot I}}+{{R_K}^I}_{LJ\perp <D>}+{C^M}_{L<D>}{{R_K}^I}_{JM}-
$$
$$
{(-)}^{\mid LJ\mid }{C^M}_{J<D>}{{R_K}^I}_{LM}-{T^M}_{JL}{P_{K\cdot
M<D>}^{\cdot I}}+
$$
$$
{P^{<A>}}_{L<D>}{P_{K\cdot J<A>}^{\cdot I}}-{(-)}^{\mid LJ\mid }{{P^{<A>}}%
_{J<D>}P}{_{K\cdot L<A>}^{\cdot I}}-{{R^{<A>}}_{JL}S}{_{K\cdot
<A><D>}^{\cdot I}}=0,
$$
$$
{{P_{<C>}}^{<A>}}_{J<D>\mid L}-{(-)}^{\mid LJ\mid }{{P_{<C>}}^{<A>}}%
_{L<D>\mid J}+{R_{<C>\cdot LJ\mid <D>}^{\cdot <A>}}+
$$
$$
{C^M}_{L<D>}{{R_{<C>}}^{<A>}}_{JM}-{(-)}^{\mid LJ\mid }{C^M}_{J<D>}{{R_{<C>}}%
^{<A>}}_{LM}-
$$
$$
{T^M}_{JL}{{P_{<C>}}^{<A>}}_{M<D>}+{P^{<F>}}_{L<D>}{{P_{<C>}}^{<A>}}_{J<F>}-
$$
$$
{(-)}^{\mid LJ\mid }{P^{<F>}}_{J<D>}{{P_{<C>}}^{<A>}}_{L<F>}-{R^{<F>}}_{JL}{{%
S_{<C>}}^{<A>}}_{F<D>}=0,
$$
$$
{P_{K\cdot J<D>\perp <C>}^{\cdot I}-(-)}^{\mid <C><D>\mid }{P_{K\cdot
J<C>\perp <D>}^{\cdot I}+{S_K}_{\ <D><C>|J}^I+}
$$
$$
{C^M}_{J<D>}{P_{K\cdot M<C>}^{\cdot I}}-{(-)}^{\mid <C><D>\mid }{C^M}_{J<C>}{%
P_{K\cdot M<D>}^{\cdot I}}+
$$
$$
{P^{<A>}}_{J<C>}{S_{K\cdot <D><A>}^{\cdot I}}-{(-)}^{\mid <C><D>\mid }{%
P^{<A>}}_{J<D>}{S_{K\cdot <C><A>}^{\cdot I}}+
$$
$$
{S^{<A>}}_{<C><D>}{P_{K\cdot J<A>}^{\cdot I}}=0,
$$
$$
{{P_{<B>}}^{<A>}}_{J<D>\perp <C>}-{(-)}^{\mid <C><D>\mid }{{P_{<B>}}^{<A>}}%
_{J<C>\perp <D>}+{{S_{<B>}}^{<A>}}_{<C><D>\mid J}+
$$
$$
{C^M}_{J<D>}{{P_{<B>}}^{<A>}}_{M<C>}-{(-)}^{\mid <C><D>\mid }{C^M}_{J<C>}{{%
P_{<B>}}^{<A>}}_{M<D>}+
$$
$$
{P^{<F>}}_{J<C>}{{S_{<B>}}^{<A>}}_{<D><F>}-{(-)}^{\mid <C><D>\mid }{P^{<F>}}%
_{J<D>}{{S_{<B>}}^{<A>}}_{<C><F>}+
$$
$$
{S^{<F>}}_{<C><D>}{{P_{<B>}}^{<A>}}_{J<F>}=0,
$$
$$
\sum\limits_{SC[<B>,<C>,<D>\}}[S_{K.<B><C>\perp
<D>}^{.I}-S_{.<B><C>}^{<A>}S_{K.<D><A>}^{.I}]=0,
$$
$$
\sum\limits_{SC[<B>,<C>,<D>\}}[S_{<F>.<B><C>\perp
<D>}^{.<A>}-S_{.<B><C>}^{<E>}S_{<F>.<E><A>}^{.<A>}]=0,
$$
where $\sum\limits_{SC[<B>,<C>,<D>\}}i$s the supersymmetric cyclic sum over
indices $<B>,$ $<C>,<D>.$

As a consequence of a corresponding arrangement of (12) we obtain the Ricci
identities (for simplicity we establish them only for ds-vector fields,
although they may be written for every ds-tensor field):
$$
{D_{[X}^{(h)}}{D_{Y\}}^{(h)}}hZ=R(hX,hY)hZ+{D_{[hX,hY\}}^{(h)}}%
hZ+\sum\limits_{f=1}^z{D}_{[hX,hY\}}^{(v_f)}hZ,\eqno(17)
$$
$$
{D}_{[X}^{(v_p)}{D_{Y\}}^{(h)}}hZ=R(v_pX,hY)hZ+{D}_{[v_pX,hY\}}^{(h)}hZ+\sum%
\limits_{f=1}^z{D}_{[v_pX,hY\}}^{(v_f)}hZ,
$$
$$
{D}_{[X}^{(v_p)}D_{Y\}}^{(v_P)}=R(v_pX,v_pY)hZ+\sum\limits_{f=1}^z{D}%
_{[v_pX,v_pY\}}^{(v_f)}hZ
$$
and
$$
{D_{[X}^{(h)}}{D_{Y\}}^{(h)}}v_pZ=R(hX,hY)v_pZ+{D_{[hX,hY\}}^{(h)}}%
v_pZ+\sum\limits_{f=1}^z{D}_{[hX,hY\}}^{(v_f)}v_pZ,\eqno(18)
$$
$$
D_{[X}^{(v_f)}D_{Y\}}^{(h)}v_pZ=R(v_fX,hY)v_pZ+\sum%
\limits_{q=1}^zD_{[v_fX,hY\}}^{(v_q)}v_pZ+\sum\limits_{q=1}^zD_{[v_fX,hY%
\}}^{(v_q)}v_pZ,
$$
$$
D_{[X}^{(v_q)}D_{Y\}}^{(v_f)}v_pZ=R(v_qX,v_fY)v_pZ+\sum%
\limits_{s=1}^zD_{[v_fX,v_fY\}}^{(v_s)}v_pZ.
$$
Putting $X={X^I}(u){\frac \delta {\delta x^I}}+{X^{<A>}}(u){\frac \delta
{\partial y^{<A>}}}$ and taking into account the local form of the h- and
v-covariant s-derivatives and (9),(10),(12),(13) we can express
respectively identities (17) and (18) in this form:
$$
{X^{<A>}}_{\mid K\mid L}-{(-)}^{\mid KL\mid }{X^{<A>}}_{\mid L\mid K}=
$$
$$
{{{R_{<B>}}^{<A>}}_{KL}}{X^{<B>}}-{T^H}_{KL}{X^{<A>}}_{\mid H}-{R^{<B>}}_{KL}%
{X^{<A>}}_{\perp <B>},
$$
$$
{X^I}_{\mid K\perp <D>}-{(-)}^{\mid K<D>\mid }{X^I}_{\perp <D>\mid K}=
$$
$$
{P_{H\cdot K<D>}^{\cdot I}}{X^H}-{C^H}_{K<D>}{X^I}_{\mid H}-{P^{<A>}}_{K<D>}{%
X^I}_{\perp <A>},
$$
$$
{X^I}_{\perp <B>\perp <C>}-{(-)}^{\mid <B><C>\mid }{X^I}_{\perp <C>\perp
<B>}=
$$
$$
{S_{H\cdot <B><C>}^{\cdot I}}{X^H}-{S^{<A>}}_{<B><C>}{X^I}_{\perp <A>}
$$
and
$$
{X^{<A>}}_{\mid K\mid L}-{(-)}^{\mid KL\mid }{X^{<A>}}_{\mid L\mid K}=
$$
$$
{{R_{<B>}}^{<A>}}_{KL}{X^{<B>}}-{T^H}_{KL}{X^{<A>}}_{\mid H}-{R^{<B>}}_{KL}{%
X^{<A>}}_{\perp <B>},
$$
$$
{X^{<A>}}_{\mid K\perp <B>}-{(-)}^{\mid <B>K\mid }{X^{<A>}}_{\perp <B>\mid
K}=
$$
$$
{{P_{<B>}}^{<A>}}_{KC}{X^C}-{C^H}_{K<B>}{X^{<A>}}_{\mid H}-{P^{<D>}}_{K<B>}{%
X^{<A>}}_{\perp <D>},
$$
$$
{X^{<A>}}_{\perp <B>\perp <C>}-{(-)}^{\mid <C><B>\mid }{X^{<A>}}_{\perp
<C>\perp <B>}=
$$
$$
{{S_{<D>}}^{<A>}}_{<B><C>}{X^{<D>}}-{S^{<D>}}_{<B><C>}{X^{<A>}}_{\perp <D>}.
$$
We note that the above presented formulas generalize for higher order
anisotropy the similar ones for locally anisotropic superspaces \cite{vlasg}.

\section{Cartan Structure Equations in DVS--Bund\-les}

Let consider a ds-tensor field on $\tilde {{\cal E}}^{<z>}$:
$$
t={t_{<A>}^I}{\delta }_I{\otimes }{\delta ^{<A>}}.
$$
The d-connection 1-forms ${\omega }_J^I$ and ${{\tilde \omega }_{<B>}^{<A>}}$
are introduced as%
$$
Dt=(D{t_{<A>}^I}){\delta }_I{\otimes }{\delta }^{<A>}
$$
with
$$
Dt_{<A>}^I=dt_{<A>}^I+{\omega }_J^I{t_{<A>}^J}-{{\tilde \omega }_{<A>}^{<B>}}%
{t_{<B>}^I}=t_{<A>\mid J}^I{dx^J}+t_{<A>\perp <B>}^I{\delta }y^{<B>}.
$$
For the d-connection 1-forms of a d-connection $D$ on $\tilde {{\cal E}%
}^{<z>}$ defined by ${{\omega }_J^I}$ and ${{\tilde \omega }_{<B>}^{<A>}}$
one holds the following structure equations:
$$
d({d^I})-{d^H}\wedge {\omega }_H^I=-{\Omega },~d{({{\delta }^{<A>}})}-{{%
\delta }^{<B>}}\wedge {{\tilde \omega }_{<B>}^{<A>}}=-{{\tilde \Omega }^{<A>}%
},
$$
$$
d{{\omega }_J^I}-{{\omega }_J^H}\wedge {{\omega }_H^I}=-{{\Omega }_J^I},~d{{%
\tilde \omega }_{<B>}^{<A>}}-{{\tilde \omega }_{<B>}^{<C>}}\wedge {{\tilde
\omega }_{<C>}^{<A>}}=-{{\tilde \Omega }_{<B>}^{<A>}},
$$
in which the torsion 2-forms ${\Omega }^I$ and ${{\tilde \Omega }^{<A>}}$
are given respectively by formulas:
$$
{{\Omega }^I}={\frac 12}{T^I}_{JK}{d^J}\wedge {d^K}+{\frac 12}{C^I}_{J<C>}{%
d^J}\wedge {{\delta }^{<C>}},
$$
$$
{{\tilde \Omega }^{<A>}}={\frac 12}{R^{<A>}}_{JK}{d^J}\wedge {d^K}+{\frac 12}%
{P^{<A>}}_{J<C>}{d^J}\wedge {{\delta }^{<C>}}+{\frac 12}{S^{<A>}}_{<B><C>}{{%
\delta }^{<B>}}\wedge {{\delta }^{<C>}},
$$
and
$$
{{\Omega }_J^I}={\frac 12}{{R_J}^I}_{KH}{d^K}\wedge {d^H}+{\frac 12P}{%
_{J\cdot K<C>}^{\cdot I}}{d^K}\wedge {{\delta }^{<C>}}+{\frac 12S}{_{J\cdot
K<C>}^{\cdot I}}{{\delta }^{<B>}}\wedge {{\delta }^{<C>}},
$$
$$
{{\tilde \Omega }_{<B>}^{<A>}}={\frac 12R}{_{<B>\cdot KH}^{\cdot <A>}}{d^K}%
\wedge {d^H}+
$$
$$
{\frac 12}{{P_{<B>}}^{<A>}}_{K<C>}{d^K}\wedge {{\delta }^{<C>}}+{\frac 12}{{%
S_{<B>}}^{<A>}}_{<C><D>}{{\delta }^{<C>}}\wedge {{\delta }^{<D>}}.
$$
We have defined the exterior product on s-space as to satisfy the property
$$
{{\delta }^{<\alpha >}}\wedge {{\delta }^{<\beta >}}=-{(-)}^{\mid <\alpha
><\beta >\mid }{{\delta }^{<\beta >}}\wedge {{\delta }^{<\alpha >}}.
$$

\section{Metrics in DVS--Bundles}

The base $\tilde M$ of dvs-bundle $\tilde {{\cal E}}^{<z>}$is considered to
be a connected and paracompact s-manifold.

A metric structure on the total space $\tilde E^{<z>}$ of a dvs-bundle $%
\tilde {{\cal E}}^{<z>}$is a supersymmetric, second order, covariant
s-tensor field
$$
G=G_{<\alpha ><\beta >}\partial ^{<\alpha >}\otimes \partial ^{<\beta >}
$$
which in every point $u\in \tilde {{\cal E}}^{<z>}$ is given by
nondegenerate supersymmetric matrix $G_{<\alpha ><\beta >}=G({{\partial }%
_{<\alpha >}},{{\partial }_{<\beta >}}){\quad }$ (with nonvanishing
superdeterminant, $sdetG\not =0).$

The metric and N-connection structures on $\tilde {{\cal E}}^{<z>}$ are
compatible if there are satisfied conditions$:$
$$
G({{\delta }_I},{{\partial }_{<A>}})=0,G(\delta _{A_f},{\partial }%
_{A_p})=0,~z\geq p>f\geq 1,
$$
or, in consequence,
$$
{G_{I<A>}}-{N_I^{<B>}}{h_{<A><B>}}=0,{G}_{A_fA_p}-{N}_{A_f}^{B_p}{h}%
_{A_pB_p}=0,\eqno(19)
$$
where
$$
{G_{I<A>}}=G({{\partial }_I},{{\partial }_{<A>}}),{G}_{A_fA_p}=G({\partial }%
_{A_f},{\partial }_{A_p}).
$$
From (42) one follows
$$
{N_I^{<B>}}={h^{<B><A>}}{G_{I<A>}},~{N}_{A_f}^{A_p}={h}^{A_pB_p}{G}%
_{A_fB_p},...,
$$
where matrices $h^{<A><B>},{h}^{A_pB_p},...$ are respectively s--inverse to
matrices
$$
h_{<A><B>}=G({{\partial }_{<A>}},{{\partial }_{<B>}}),h_{A_pB_p}=G({\partial
}_{A_p},{\partial }_{B_p}).
$$
So, in this case, the coefficients of N-connection are uniquely determined
by the components of the metric on $\tilde {{\cal E}}^{<z>}.$

A compatible with N--connection metric on $\tilde {{\cal E}}^{<z>}$ is
written in irreducible form as
$$
G(X,Y)=G(hX,hY)+G(v_1X,v_1Y)+...+G(v_zX,v_zY),{\quad }X,Y\in {\Xi (\tilde {%
{\cal E}}^{<z>})},
$$
and looks locally as
$$
G=g_{{\alpha }{\beta }}{(u)}{{\delta }^\alpha }\otimes {{\delta }^\beta }%
=g_{IJ}{d^I}\otimes {d^J}+h_{<A><B>}{{\delta }^{<A>}}\otimes {{\delta }^{<B>}%
}=
$$
$$
g_{IJ}{d^I}\otimes {d^J}+h_{A_1B_1}{\delta }^{A_1}\otimes {\delta }%
^{B_1}+h_{A_2B_2}{\delta }^{A_2}\otimes {\delta }^{B_2}+...+h_{A_zB_z}{%
\delta }^{A_z}\otimes {\delta }^{B_z}.\eqno(20)
$$

A d-connection $D$ on $\tilde {{\cal E}}^{<z>}$ is metric, or compatible
with metric $G$, if conditions
$$
{D_{<\alpha >}}{G_{<{\beta ><}{\gamma >}}}=0
$$
are satisfied.

A d-connection $D$ on $\tilde {{\cal E}}^{<z>}$ provided with a metric $G$
is a metric d-connection if and only if
$$
{D_X^{(h)}}{(hG)}=0,{D_X^{(h)}}{(v}_p{G)}=0,{D}_X^{(v_p)}{(hG)}=0,{D}%
_X^{(v_f)}{(v}_p{G)}=0\eqno(21)
$$
for every $,f,p=1,2,...,z,$ and $X\in {\Xi (\tilde {{\cal E}}^{<z>})}.$
Conditions (21) are written in locally adapted form as
$$
g_{IJ\mid K}=0,g_{IJ\perp <A>}=0,h_{<A><B>\mid K}=0,h_{<A><B>\perp <C>}=0.
$$

In every dvs--bundle provided with compatible N--connection and metric
structures one exists a metric d-connection (called the canonical
d-connection associated to $G)$ depending only on components of G-metric and
N-connection. Its local coefficients $C{\Gamma }=({{\grave L}^I}_{JK},{{%
\grave L}^{<A>}}_{<B>K},{{\grave C}^I}_{J<C>},{{\grave C}^{<A>}}_{<B><C>})$
are as follows:
$$
{{\grave L}^I}_{JK}={\frac 12}{g^{IH}}({{{\delta }_K}g_{HJ}+{{\delta }_J}%
g_{HK}-{{\delta }_H}g_{JK}}),\eqno(22)
$$
$$
{{\grave L}^{<A>}}_{<B>K}={\delta _{<B>}}{N_K^{<A>}}+
$$
$$
{\frac 12}{h^{<A><C>}}[{{{\delta }_{<K>}}{h_{<B><C>}}-(\delta }%
_{<B>}N_K^{<D>})h_{<\dot D><C>}{-{(\delta }_{<C>}N_K^{<D>})h_{<\dot D><B>}}%
],
$$
$$
{{\grave C}^I}_{J<C>}={\frac 12}{g^{IK}\delta }{_{<C>}}{g_{JK}},
$$
$$
{{\grave C}^{<A>}}_{<B><C>}={\frac 12}{h^{<A><D>}(\delta }_{<C>}h_{<D><B>}+{%
\delta }_{<B>}h_{<D><C>}-{\delta }_{<D>}h_{<B><C>}.
$$
We emphasize that, in general, the torsion of $C\Gamma $--connection (22)
does not vanish.

It should be noted here that on dvs-bundles provided with N-connection and
d-connection and metric really it is defined a multiconnection ds-structure,
i.e. we can use in an equivalent geometric manner different types of d-
connections with various properties. For example, for modeling of some
physical processes we can use an extension of the Berwald d--connection
$$
B{\Gamma }=({{L^I}_{JK}},\delta _{<B>}N_K^{<A>},0,{C}_{<B><C>}^{<A>}),%
\eqno(23)
$$
where ${L^I}_{JK}={{\grave L}^I}_{JK}$ and ${C^{<A>}}_{<B><C>}={{\grave C}%
^{<A>}}_{<B><C>},$ which is hv-metric, i.e. satisfies conditions:
$$
{D_X^{(h)}}hG=0,...,{D}_X^{(v_p)}v_pG=0,...,{D}_X^{(v_z)}v_zG=0
$$
for every $X\in \Xi {(\tilde {{\cal E}}^{<z>})},$ or in locally adapted
coordinates,
$$
g_{IJ\mid K}=0,h_{<A><B>\perp <C>}=0.
$$

As well we can introduce the Levi-Civita connection%
$$
\{{\frac{<{\alpha >}}{<{{\beta ><}{\gamma >}}}}\}={\frac 12}{G^{<{\alpha ><}{%
\beta >}}({{\partial }_{<\beta >}}{G_{<\tau ><\gamma >}}+{{\partial }%
_{<\gamma >}}{G_{<\tau ><\beta >}}-{{\partial }_{<\tau >}}{G_{<\beta
><\gamma >}})},
$$
constructed as in the Riemann geometry from components of metric $G_{<{%
\alpha ><}{\beta >}}$ by using partial derivations ${{\partial }_{<\alpha >}}%
={\frac \partial {\partial u^{<\alpha >}}}=({\frac \partial {\partial x^I}},{%
\frac \partial {\partial y^{<A>}}}),$ which is metric but not a d-connection.

In our further considerations we shall largely use the Christoffel
d--symbols defined similarly as components of Levi-Civita connection but by
using la-partial derivations,
$$
{{{\tilde \Gamma }^{<\alpha >}}_{<\beta ><\gamma >}}={\frac 12}{G^{<\alpha
><\tau >}}({{\delta }_{<\beta >}}{G_{<\tau ><\gamma >}}+{{\delta }_{<\gamma
>}}{G_{<\tau ><\beta >}}-{{\delta }_{<\tau >}}{G_{<\beta ><\gamma >}}),%
\eqno(24)
$$
having components
$$
C{\tilde \Gamma }=({L^I}_{JK},0,0,{C^{<A>}}_{<B><C>}),
$$
where coefficients ${L^I}_{JK}$ and ${C^{<A>}}_{<B><C>}$ must be com\-puted
as in formu\-las (20).

We can express arbitrary d-connection as a deformation of the background
d-connection (23):
$$
{{{\Gamma }^{<\alpha >}}_{<\beta ><\gamma >}}={{\tilde \Gamma }_{\cdot
<\beta ><\gamma >}^{<\alpha >}}+{{P^{<\alpha >}}_{<\beta ><\gamma >}},%
\eqno(25)
$$
where ${{P^{<\alpha >}}_{<\beta ><\gamma >}}$ is called the deformation
ds-tensor. Putting splitting (25) into (19) and (23) we can express torsion $%
{T^{<\alpha >}}_{<\beta ><\gamma >}$ and curvature\\ ${{R_{<\beta >}}%
^{<\alpha >}}_{<\gamma ><\delta >}$ of a d-connection ${{\Gamma }^{<\alpha >}%
}_{<\beta ><\gamma >}$ as respective deformations of torsion ${{\tilde T}%
^{<\alpha >}}_{<\beta ><\gamma >}$ and torsion ${\tilde R}_{<\beta >\cdot
<\gamma ><\delta >}^{\cdot <\alpha >}$ for connection ${{\tilde \Gamma }%
^{<\alpha >}}_{<\beta ><\gamma >}{\quad }:$
$$
{{T^{<\alpha >}}_{<\beta ><\gamma >}}={{\tilde T}_{\cdot <\beta ><\gamma
>}^{<\alpha >}}+{{\ddot T}_{\cdot <\beta ><\gamma >}^{<\alpha >}}
$$
and
$$
{{{R_{<\beta >}}^{<\alpha >}}_{<\gamma ><\delta >}}={{\tilde R}_{<\beta
>\cdot <\gamma ><\delta >}^{\cdot <\alpha >}}+{{\ddot R}_{<\beta >\cdot
<\gamma ><\delta >}^{\cdot <\alpha >}},
$$
where
$$
{{\tilde T}^{<\alpha >}}_{<\beta ><\gamma >}={{\tilde \Gamma }^{<\alpha >}}%
_{<\beta ><\gamma >}-{(-)}^{\mid <\beta ><\gamma >\mid }{{\tilde \Gamma }%
^{<\alpha >}}_{<\gamma ><\beta >}+{w^{<\alpha >}}_{<\gamma ><\delta >},
$$
$$
~{{\ddot T}^{<\alpha >}}_{<\beta ><\gamma >}={{\ddot \Gamma }^{<\alpha >}}%
_{<\beta ><\gamma >}-{(-)}^{\mid <\beta ><\gamma >\mid }{{\ddot \Gamma }%
^{<\alpha >}}_{<\gamma ><\beta >},
$$
and%
$$
{{\tilde R}_{<\beta >\cdot <\gamma ><\delta >}^{\cdot <\alpha >}}={{\delta }%
_{<\delta >}}{{\tilde \Gamma }^{<\alpha >}}_{<\beta ><\gamma >}-{(-)}^{\mid
<\gamma ><\delta >\mid }{{\delta }_{<\gamma >}}{{\tilde \Gamma }^{<\alpha >}}%
_{<\beta ><\delta >}+
$$
$$
{{{\tilde \Gamma }^{<\varphi >}}_{<\beta ><\gamma >}}{{{\tilde \Gamma }%
^{<\alpha >}}_{<\varphi ><\delta >}}-{(-)}^{\mid <\gamma ><\delta >\mid }{{{%
\tilde \Gamma }^{<\varphi >}}_{<\beta ><\delta >}}{{{\tilde \Gamma }%
^{<\alpha >}}_{<\varphi ><\gamma >}}+{{\tilde \Gamma }^{<\alpha >}}_{<\beta
><\varphi >}{w^{<\varphi >}}_{<\gamma ><\delta >},
$$
$$
{{\ddot R}_{<\beta >\cdot <\gamma ><\delta >}^{\cdot <\alpha >}}={{\tilde D}%
_{<\delta >}}{{P^{<\alpha >}}_{<\beta ><\gamma >}}-{(-)}^{\mid <\gamma
><\delta >\mid }{{\tilde D}_{<\gamma >}}{{P^{<\alpha >}}_{<\beta ><\delta >}}%
+
$$
$$
{{P^{<\varphi >}}_{<\beta ><\gamma >}}{{P^{<\alpha >}}_{<\varphi ><\delta >}}%
-{(-)}^{\mid <\gamma ><\delta >\mid }{{P^{<\varphi >}}_{<\beta ><\delta >}}{{%
P^{<\alpha >}}_{<\varphi ><\gamma >}}
$$
$$
+{{P^{<\alpha >}}_{<\beta ><\varphi >}}{{w^{<\varphi >}}_{<\gamma ><\delta >}%
},
$$
the nonholonomy coefficients ${w^{<\alpha >}}_{<\beta ><\gamma >}$ are
defined as
$$
[{\delta }_{<\alpha >},{\delta }_{<\beta >}\}={{\delta }_{<\alpha >}}{{%
\delta }_{<\beta >}}-{(-)}^{|<\alpha ><\beta >|}{{\delta }_{<\beta >}}{{%
\delta }_{<\alpha >}}={w^{<\tau >}}_{<\alpha ><\beta >}{{\delta }_{<\tau >}}.
$$

We emphasize that if from geometric point of view all considered
d--con\-nect\-i\-ons are ''equ\-al in rights'', the con\-struct\-ion of
physical models on la-spaces requires an explicit fixing of the type of
d--con\-nect\-ion and metric structures.

\section{Higher Order Tangent S--Bundles}

The aim of this section is to present a study of supersymmetric extensions
from $\widetilde{M}$ to $T\tilde M$ and $Osc^{(z)}\widetilde{M}$ and to
consider corresponding prolongations of Riemann and generalized Finsler
structures (on classical and new approaches to Finsler geometry, its
generalizations and applications in physics se, for example, \cite
{fin,car,run,ma87,ma94,asa,asa88,mat,mk,am,az94,bog,bej}).

The presented in the previous sections basic results on dvs-bundles ${\tilde {%
{\cal E}}^{<z>}}$ pro\-vid\-ed with N-connection, d-connection and metric
structures can be correspondingly adapted to the osculator s--bundle $\left(
Osc^z\tilde M,\pi ,\tilde M\right) .$ In this case the dimension of the base
space and typical higher orders fibre coincides and we shall not distinguish
indices of geometrical objects.

Coefficients of a d--connection $D\Gamma
(N)=(L_{JM}^I,C_{(1)JM}^I,...,C_{(z)JM}^I)$ in $Osc^z\tilde M,$ with respect
to a la--base are introduced as to satisfy equations%
$$
D_{\frac \delta {\delta x^I}}\frac \delta {\delta y_{(f)}^I}=L_{IJ}^M\frac
\delta {\delta y_{(f)}^M},~D_{\frac \delta {\delta y_{(p)}^J}}\frac \delta
{\delta y_{(f)}^I}=C_{(p)IJ}^M\frac \delta {\delta y_{(f)}^M},\eqno(26)
$$
$$
(f=0,1,...,z;p=1,...,z,\mbox{ and }~y_{(0)}^I=x^I).
$$

A metric structure on $Osc^z\tilde M$ is ds-tensor s--symmetric field $%
g_{IJ}(u_{(z)})=g_{IJ}(x,y_{(1)},y_{(2)},...,y_{(z)})$ of type $%
(0,2),srank|g_{ij}|=(n,m).$ The N--lift of Sasaki type of $g_{IJ}$ is given
by (see (20)) defines a global Riemannian s--structure (if $\widetilde{M}$
is a s--differentiable, paracompact s-manifold):%
$$
G=g_{IJ}(u_{(z)})dx^I\otimes dx^J+g_{IJ}(u_{(z)})dy_{(1)}^I\otimes
dy_{(1)}^J+...+g_{IJ}(u_{(z)})dy_{(z)}^I\otimes dy_{(z)}^J.\eqno(27)
$$
The condition of compatibility of a d--connection (26) with metric (27) is
expressed as
$$
D_XG=0,\forall X\in \Xi (Osc^z\tilde M),
$$
or, by using d--covariant partial derivations $|_{(p)}$ defined by
coefficients\\ $(L_{JM}^I,C_{(1)JM}^I,...,C_{(z)JM}^I),$
$$
g_{IJ|M=0,~}g_{IJ|_{(p)M}}=0,(p=1,...,z).
$$
An example of compatible with metric d--connection is given by Christoffel
d-symbols (see (24)):%
$$
L_{IJ}^M=\frac 12g^{MK}\left( \frac{\delta g_{KJ}}{\partial x^I}+\frac{%
\delta g_{IK}}{\partial x^J}-\frac{\delta g_{IJ}}{\partial x^K}\right) ,%
$$
$$
C_{(p)IJ}^M=\frac 12g^{MK}\left( \frac{\delta g_{KJ}}{\partial y_{(p)}^I}+%
\frac{\delta g_{IK}}{\partial y_{(p)}^J}-\frac{\delta g_{IJ}}{\partial
y_{(p)}^K}\right) ;p=1,2,...,z.
$$

\section{Supersymmetric Extensions of Finsler Spa\-ces}

We start our considerations with the ts-bundle $T\tilde M.$ An s-vector $%
X\in \Xi (T\tilde M)$ is decomposed with respect to la--bundles as%
$$
X=X(u)^I\delta _I+Y(u)^I\partial _I,
$$
where $u=u^\alpha =(x^I,y^J)$ local coordinates. The s--tangent structures
(6) are transformed into a global map
$$
J:\Xi (T\tilde M)\to \Xi (T\tilde M)
$$
which does not depend on N-connection structure:
$$
J({\frac \delta {\delta x^I}})={\frac \partial {\partial y^I}}
$$
and%
$$
J({\frac \partial {\partial y^I}})=0.
$$
This endomorphism is called the natural (or canonical) almost tangent
structure on $T\tilde M;$ it has the properties:
$$
1)J^2=0,{\quad }2)ImJ=KerJ=VT\tilde M
$$
and 3) the Nigenhuis s-tensor,
$$
{N_J}(X,Y)=[JX,JY\}-J[JX,Y\}-J[X,JY]
$$
$$
(X,Y\in \Xi (TN))
$$
identically vanishes, i.e. the natural almost tangent structure $J$ on $%
T\tilde M$ is integrable.

A generalized Lagrange superspace, GLS-- space, is a pair\\ ${GL}%
^{n,m}=(\tilde M,g_{IJ}(x,y))$, \quad where \quad $g_{IJ}(x,y)$ \quad is a
ds-- tensor field on \\ \quad ${{\tilde {T{\tilde M}}}=T\tilde M-\{0\}},$%
\quad s--symmetric of superrank $(n,m).$

We call $g_{IJ}$ as the fundamental ds-tensor, or metric ds-tensor, of
GLS-space.

There exists an unique d-connection $C\Gamma (N)$ which is compatible with $%
g_{IJ}{(u)}$ and has vanishing torsions ${T^I}_{JK}$ and ${S^I}_{JK}$ (see
formulas (26) rewritten for ts-bundles). This connection, depending only on $%
g_{IJ}{(u)}$ and ${N_J^I}{(u)}$ is called the canonical metric d-connection
of GLS-space. It has coefficients
$$
{L^I}_{JK}={\frac 12}{g^{IH}}({\delta }_J{g_{HK}}+{\delta }_H{g_{JK}}-{%
\delta }_H{g_{JK}}),
$$
$$
{C^I}_{JK}={\frac 12}{g^{IH}}({\partial }_J{g_{HK}}+{\partial }_H{g_{JK}}-{%
\partial }_H{g_{JK}}).
$$
There is a unique normal d-connection $D\Gamma (N)=({\bar L}_{\cdot JK}^I,{%
\bar C}_{\cdot JK}^I)$ which is metric and has a priori given torsions ${T^I}%
_{JK}$ and ${S^I}_{JK}.$ The coefficients of $D\Gamma (N)$ are the following
ones:
$$
{\bar L}_{\cdot JK}^I={L^I}_{JK}-\frac 12g^{IH}(g_{JR}{T^R}_{HK}+g_{KR}{T^R}%
_{HJ}-g_{HR}{T^R}_{KJ}),
$$
$$
{\bar C}_{\cdot JK}^I={C^I}_{JK}-\frac 12g^{IH}(g_{JR}{S^R}_{HK}+g_{KR}{S^R}%
_{HJ}-g_{HR}{S^R}_{KJ}),
$$
where ${L^I}_{JK}$ and ${C^I}_{JK}$ are the same as for the $C\Gamma (N)$%
--connection (26).

The Lagrange spaces were introduced \cite{ker} in order to geometrize the
concept of Lagrangian in mechanics (the Lagrange geometry is studied in
details in \cite{ma87,ma94}). For s-spaces we present this generalization:

A Lagrange s-space, LS-space, $L^{n,m}=(\tilde M,g_{IJ}),$ is defined as a
particular case of GLS-space when the ds-metric on $\tilde M$ can be
expressed as
$$
g_{IJ}{(u)}={\frac 12}{\frac{{\partial }^2L}{{{\partial y^I}{\partial y^J}}}}%
,\eqno(28)
$$
where $L:T\tilde M\to \Lambda ,$ is a s-differentiable function called a
s-Lagrangian on $\tilde M.$

Now we consider the supersymmetric extension of the Finsler space:

A Finsler s-metric on $\tilde M$ is a function $F_S:T\tilde M\to \Lambda $
having the properties:

1. The restriction of $F_S$ to ${\tilde {T\tilde M}}=T\tilde M\setminus
\{0\} $ is of the class $G^\infty $ and F is only supersmooth on the image
of the null cross--section in the ts-bundle to $\tilde M.$

2. The restriction of F to ${\tilde {T\tilde M}}$ is positively homogeneous
of degree 1 with respect to ${(y^I)}$, i.e. $F(x,{\lambda }y)={\lambda }%
F(x,y),$ where ${\lambda }$ is a real positive number.

3. The restriction of F to the even subspace of $\tilde {T\tilde M}$ is a
positive function.

4. The quadratic form on ${\Lambda }^{n,m}$ with the coefficients
$$
g_{IJ}{(u)}={\frac 12}{\frac{{\partial }^2F^2}{{{\partial y^I}{\partial y^J}}%
}}
$$
defined on $\tilde {T\tilde M}$ is nondegenerate.

A pair $F^{n,m}=(\tilde M,F)$ which consists from a supersmooth
s-ma\-ni\-fold $\tilde M$ and a Finsler s-metric is called a Finsler
superspace, FS-space.

It's obvious that FS-spaces form a particular class of LS-spaces with
s-Lagran\-gi\-an $L={F^2}$ and a particular class of GLS-spaces with metrics
of type (28).

For a FS-space we can introduce the supersymmetric variant of nonlinear
Cartan connection \cite{car,run} :
$$
N_J^I{(x,y)}={\frac \partial {\partial y^J}}G^{*I},
$$
where
$$
G^{*I}={\frac 14}g^{*IJ}({\frac{{\partial }^2{\varepsilon }}{{\partial y^I}{%
\partial x^K}}}{y^K}-{\frac{\partial {\varepsilon }}{\partial x^J}}),{\quad }%
{\varepsilon }{(u)}=g_{IJ}{(u)}y^Iy^J,
$$
and $g^{*IJ}$ is inverse to $g_{IJ}^{*}{(u)}={\frac 12}{\frac{{\partial }%
^2\varepsilon }{{{\partial y^I}{\partial y^J}}}}.$ In this case the
coefficients of canonical metric d-connection (26) gives the supersymmetric
variants of coefficients of the Cartan connection of Finsler spaces. A
similar remark applies to the Lagrange superspaces.

\section{Higher Order Prolongations of Fins\-ler and La\-gran\-ge S--Spa\-ces
}

The geometric constructions on $T\widetilde{M}$ from the previous subsection
have corresponding generalizations to the $Osc^{(z)}\widetilde{M}$
s--bundle. The basic idea is similar to that used for prolongations of
geometric structures (see \cite{morim} for prolongations on tangent bundle).
Having defined a metric structure $g_{IJ}(x)$ on a s-manifold $\widetilde{M}$
we can extend it to the $Osc^z\tilde M$ s--bundle by considering $%
g_{IJ}(u_{(z)})=g_{IJ}(x)$ in (27). R. Miron and Gh. Atanasiu \cite{mirata}
solved the problem of prolongations of Finsler and Lagrange structures on
osculator bundle. In this subsection we shall analyze supersymmetric
extensions of Finsler and Lagrange structures as well present a brief
introduction into geometry of higher order Lagrange s-spaces.

Let $F^{n,m}=(\tilde M,F)$ be a FS--space with the fundamental function $%
F_S:T\tilde M\to \Lambda $ on $\widetilde{M}.$ A prolongation of $F$ on $%
Osc^z\tilde M$ is given by a map%
$$
(F\circ \pi _1^z)(u_{(z)})=F(u_{(1)})
$$
and corresponding fundamental tensor
$$
g_{IJ}(u_{(1)})=\frac 12\frac{\partial ^2F^2}{\partial y_{(1)}^I\partial
y_{(1)}^J},
$$
for which
$$
(g_{IJ}\circ \pi _1^z)(u_{(z)})=g_{IJ}(u_{(1)}).
$$
So, $g_{IJ}(u_{(1)})$ is a ds--tensor on\\ $\widetilde{Osk^z\widetilde{M}}%
=Osc^z\tilde M/\{0\}=\{(u_{(z)})\in Osc^z\tilde M,srank|y_{(1)}^I|=1\}.$

The Christoffel symbols
$$
\gamma _{IJ}^M(u^{(1)})=\frac 12g^{MK}(u_{(1)})(\frac{\partial
g_{KI}(u_{(1)})}{\partial x^J}+\frac{\partial g_{JK}(u_{(1)})}{\partial x^I}-%
\frac{\partial g_{IJ}(u_{(1)})}{\partial x^K})
$$
define the Cartan nonlinear connection \cite{car35}:%
$$
G_{(N)J}^I=\frac 12\frac \partial {\partial y_{(1)}^J}(\gamma
_{KM}^Iy_{(1)}^Ky_{(1)}^M).\eqno(29)
$$
The dual coefficients for the N-connection (see formulas (5)) are
recurrently computed by using (29) and operator
$$
\Gamma =y_{(1)}^I\frac \partial {\partial x^I}+2~y_{(2)}^I\frac \partial
{\partial y_{(1)}^I}+...+z~y_{(z)}^I\frac \partial {\partial y_{(z-1)}^I},
$$
$$
M_{(1)J}^I=G_{(N)J}^I,
$$
$$
M_{(2)J}^I=\frac 12[\Gamma G_{(N)J}^I+G_{(N)K}^IM_{(1)J}^K],
$$
$$
..............
$$
$$
M_{(z)J}^I=\frac 1z[\Gamma M_{(z-1)J}^I+G_{(N)K}^IM_{(z-1)J}^K].
$$

The prolongations of FS--spaces can be generalized for Lagrange s--spaces
(on Lagrange spaces and theirs higher order extensions see \cite
{ma87,ma94,mirata} and on supersymmetric extensions of Finsler geometry see
\cite{vlasg}). Let $L^{n,m}=(\tilde M,g_{IJ})$ be a Lagrange s--space. The
Lagrangian $L:T\widetilde{M}\rightarrow \Lambda \,$ can be extended on $%
Osc^z\tilde M$ by using maps of the Lagrangian, $(L\circ \pi
_1^z)(u_{(z)})=L\left( u_{(1)}\right) ,$ and, as a consequence, of the
fundamental tensor (28), $(g_{IJ}\circ \pi _1^z)(u_{(z)})=g_{IJ}\left(
u_{(1)}\right) .$


We introduce the notion of Lagrangian of z--order on a differentiable
s--manifold $\widetilde{M}$ as a map $L^z:Osc^z\tilde M$ $\rightarrow
\Lambda .$ In order to have concordance with the definitions proposed by
\cite{mirata} we require the even part of the fundamental ds--tensor to be
of constant signature. Here we also note that questions to considered in
this subsection, being an supersymmetric approach, are connected with the
problem of elaboration of the so--called higher order analytic mechanics
(see, for instance, \cite{cram,lib,leo,sau}).

A Lagrangian s--differentiable of order $z$ ($z=1,2,3,...)$ on
s-differentiable s--manifold $\widetilde{M}$ is an application $L^{(z)}:Osc^z%
\widetilde{M}\rightarrow \Lambda ,$ s--differentiable on $\widetilde{Osk^z%
\widetilde{M}}$ and smooth in the points of $Osc^z\widetilde{M}$ where $%
y_{(1)}^I=0.$

It is obvious that
$$
g_{IJ}(x,y_{(1)},...,y_{(z)})=\frac 12\frac{\partial ^2L^{(z)}}{\partial
y_{(z)}^I\partial y_{(z)}^J}
$$
is a ds--tensor field because with respect to coordinate transforms (1) one
holds transforms%
$$
K_I^{I^{\prime }}K_J^{J^{\prime }}g_{I^{\prime }J^{\prime }}=g_{IJ.}
$$

A Lagrangian $L$ is regular if $srank|g_{IJ}|=(n,m).$

A Lagrange s--space of $z$--order is a pair $L^{(z,n,m)}=(\widetilde{M}%
,L^{(z)}),$ where $L^{(z)}$ is a s--differentiable regular Lagrangian of $z$%
--order, and with ds--tensor $g_{IJ}$ being of constant signature on the
even part of the basic s--manifold.

For details on nonsupersymmetric osculator bundles see  \cite{mirata}.

\section{Conclusions and Discussion}
In the present work we focused on the  definition and calculation
 of basic geometric structures on superspaces  with higher order generic
 anisotropy. In the framework of such spaces it seems more convenient
 to touch on the problem of formulation of higher dimensional and locally
 anisotropic  classical and quantum field theories. It should be noted here
 that the general approach of modeling of locally anisotropic geometries
 (Finsler, Lagrange and various higher order extensions and prolongations;
 we are very much inspired by R. Miron, M. Anastasiei and G. Atanasiu works
 \cite{ma87,ma94,mirata}) made apparent the possibility and manner of
 formulation of physical theories  by adapting
 geometric and physical constructions to the N--connection structure. Former
 considerations based on straightforward applications of  Finsler
 spaces are characterized by cumbersome tensorial calculations and different
 ambiguities in physical interpretation. In our case we are dealing with
 a standard (super)bundle technique which allow us a geometric study being
 very similar to that for supersymmetic extensions of Einstein--Cartan spaces
 with torsion but additionally provided with a N--connection structure. From the
 viewpoint of modern Kaluza--Klain theories the N--connection can be considered
 as a "reduction" field from higher dimensions to lower dimensional ones.
 Higher order anisotropies can be treated in this case as relic interactions
 and (classic or quantum) fluctuations from higher dimensions. This is a matter
 of our further investigations (see also our previous works
 \cite{vjmp,vlasg,vg}).

\vskip5pt
{\bf Acknowledgment}
The author  is indebted to R Miron for support and useful discussions.

\newpage
\footnotesize{

}
\end{document}